\documentclass[aps,prb,reprint,superscriptaddress,citeautoscript]{revtex4-1}
\usepackage{graphicx}
\usepackage{dcolumn}
\usepackage{bm}
\usepackage{epstopdf}
\usepackage{amsmath}
\begin{document}
\title{Tuning biexciton binding and anti-binding in core/shell quantum dots}
\author{Peter G McDonald}
\affiliation {School of Engineering and Physical Sciences, SUPA, Heriot-Watt University, Edinburgh EH14 4AS, UK}
\author{Edward J Tyrrell}
\affiliation{Department of Materials, University of Oxford, Parks Road, Oxford OX1 3PH, UK}
\author{John Shumway}
\affiliation{Department of Physics, Arizona State University, Tempe, AZ 85287, USA}
\author{Jason M Smith}  
\affiliation{Department of Materials, University of Oxford, Parks Road, Oxford OX1 3PH, UK}
\author{Ian Galbraith} 
\email[]{i.galbraith@hw.ac.uk}
\affiliation{School of Engineering and Physical Sciences, SUPA, Heriot-Watt University, Edinburgh EH14 4AS, UK}
\date{\today}
\begin{abstract}
We use a path integral quantum Monte Carlo method to simulate excitons and biexcitons in core shell nanocrystals with Type-I, II and quasi-Type II band alignments. Quantum Monte Carlo techniques allow for all quantum correlations to be included when determining the thermal ground state, thus producing accurate predictions of biexciton binding.
These subtle quantum correlations are found to cause the biexciton to be binding with Type-I carrier localization and strongly anti-binding with Type-II carrier localization, in agreement with experiment for both core shell nanocrystals and dot in rod nanocrystal structures. Simple treatments based on perturbative approaches are shown to miss this important transition in the biexciton binding.
Understanding these correlations offers prospects to engineer strong biexciton anti-binding which is crucial to the design of nanocrystals for single exciton lasing applications.
\end{abstract}
\maketitle
\section{Introduction}
\begin{figure}[tbp]
	\centering	
\includegraphics{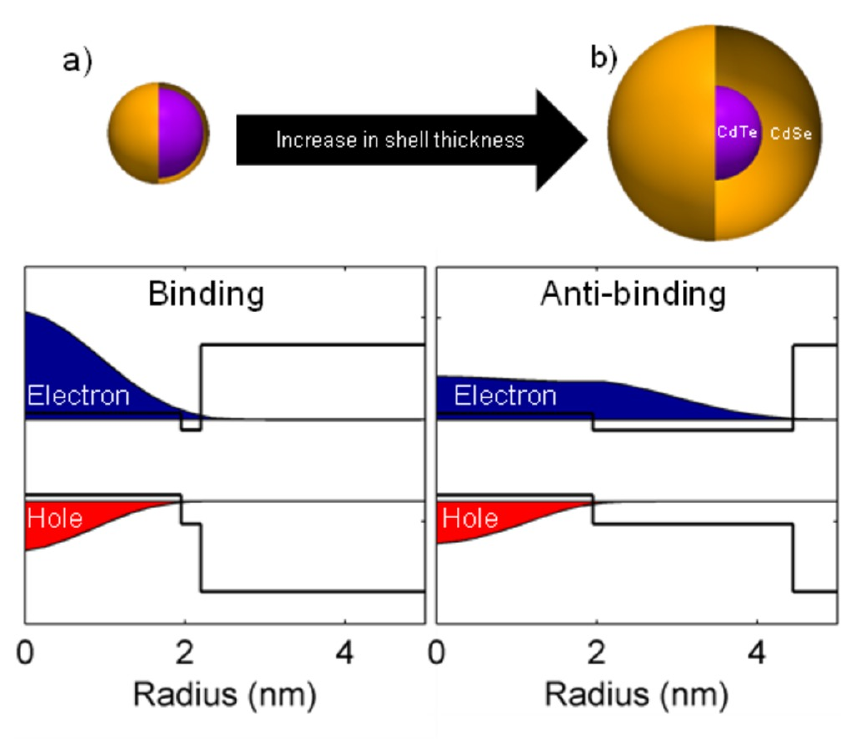}
	\caption{CdTe/CdSe Type-II core/shell nano-crystal schematic and band edges, with electron and hole probability densities within a biexciton.  a) 1.95\,nm core radius and 0.25\,nm shell thickness,  b) 1.95\,nm core radius and 2.5\,nm shell thickness. }
	\label{fig:cdtecdse_band}
\end{figure}

The properties of excitons in semiconductor quantum dots have in recent years been the focus of considerable study, both as a source of new fundamental physics and as a quantum electronic system that can be engineered for a range of applications including optical sources and detectors.
A striking example of such engineered quantum confinement is found in Type-II core-shell nanocrystals \cite{Kim03,Ivanov04,Klimov07,Oron07} as depicted in Fig.~\ref{fig:cdtecdse_band}. Exploiting independent control of the electron and hole wavefunctions it is possible to tune the exciton energies, inter-particle Coulomb and exchange interactions, and transition lifetimes to a far greater degree than in Type-I band-aligned structures. Of particular practical significance is the ability to generate positive exciton-exciton (X-X) interaction energies (anti-binding) as a result of such spatial manipulation of electrons and holes.  Then the photon energy required to generate a biexciton in a quantum dot already containing an exciton is greater than the exciton recombination energy. This is an important pre-requisite for achieving lasing in the single exciton regime since an incident photon with energy resonant with the exciton energy  may stimulate emission but cannot be absorbed. Exciton-exciton interaction energies of up to +110\,meV have been reported in CdS/ZnSe core-shell structures \cite{Klimov07}, but single exciton lasing from nanocrystals has yet to be achieved. Recently `dot-in-rod' structures in which an approximately spherical CdSe nanocrystal is embedded inside a CdS rod have also been shown to possess controllable quasi-Type II behavior \cite{Talapin04, Carbone07, Talapin07, Sitt09, Luo10, Raino11, She11}, and thus are candidate systems for single exciton lasing.

Previous calculations of the exciton-exciton interaction energy ($\Delta_{xx}=E_{xx}-2E_x$ where $E_x$ and $E_{xx}$ are the exciton and biexciton total energies) in Type- II nanocrystals have mostly relied on first order perturbation theory and  assumed spherical symmetry \cite{Klimov07, Piryatinski07, Oron07, Deutsch11}. These calculations implicitly assume the limit of strong confinement, in which the energetic separation of the single particle states is much greater than the inter-particle interaction energy. They provide a first approximation to the increase in carrier repulsion upon growth of a Type-II aligned shell layer, but consistently overestimate the repulsive effect since they do not include any spatial correlations in the carrier wavefunctions. Measurements of $\Delta_{xx}$ show that it is always negative (binding) for core-only nanocrystals, and becomes positive (anti-binding) only as the shell thickness is increased beyond a threshold value \cite{Oron07}.
Perturbative models miss this important transition between binding and anti-binding as the nanocrystal confinement passes from the Type-I to the Type-II regime.  
 Korkusinski \textit{et al}.\  employed a configuration interaction (CI) approach using a tight binding basis set to calculate the exciton and biexciton binding energies in wurtzite core-only CdSe nanocrystals \cite{Korkusinski10}, finding that biexciton anti-binding occurs for nanocrystals smaller than about 4\,nm since the crystal field provides greater localization of the quantum confined hole than the electron. In many situations CI calculations converge only slowly as the basis size is increased. This is because the influence of higher lying states drops off as $1/\Delta E$ where $\Delta E$ is the energy of the basis state relative to the system energy, and the density of states tends to increase rapidly as the radius increases and as higher energy states are included. These two trends combined make it computationally challenging to include sufficient states to be confident of good convergence.

In this work we report on path integral quantum Monte Carlo (PI-QMC) calculations of exciton binding energies and exciton-exciton interaction energies, with a full treatment of the quantum correlations in the many body system. PI-QMC does not suffer from the convergence problems of CI calculations mentioned above. 
The QMC approach has been used previously to calculate the properties of simple nanocrystals \cite{Shumway01}, the results of which illustrated the importance of correlation effects in determining the biexciton binding, but have yet to be applied to anti-binding scenarios and to  multishell heterostructures. 
Unlike perturbative approaches, our results capture the experimentally observed transition from binding to anti-binding  character in Type-II nanocrystals.   
We use this to study the Type-II regime in CdTe/CdSe and CdS/ZnSe core/shell nanocrystals, and make comparisons to previous experiemental results. We then invetigate an inverted Type-I core/shell structure, 
ZnSe/CdSe, and a quasi-Type II dot-in-rod structure, CdSe/CdS, both of which have been suggested as possible candidates for strong anti-binding.

\section{Model} 
We focus on the binding/anti-binding transition of the biexciton where an 
accurate treatment of the correlation energy is crucial to obtain the correct
magnitude and sign of the biexciton binding~\cite{Shumway01}.
As discussed in Ref.~\onlinecite{Shumway01}, we choose a simplified single-band
effective mass model to enable essentially exact determination of correlation energy
in the binding transition, while losing some of the details of atomistic models.
We model the nanostructures using a biexciton Hamiltonian of the form,
\begin{equation}
H^{XX} = H_{\text{kin}} + V_{\text{coul}} + V_{\text{dot}},
\end{equation}
where the kinetic energy arises from parabolic bands,
\begin{equation}\label{eq:kinetic}
H_{\text{kin}} =
\frac{\textbf{p}_{e_1}^2}{2m^*_{e}}
+\frac{\textbf{p}_{e_2}^2}{2m^*_{e}}
+\frac{\textbf{p}_{h_1}^2}{2m^*_{h}}
+\frac{\textbf{p}_{h_2}^2}{2m^*_{h}}.
\end{equation}
The interaction potential includes all pair-wise Coulomb interactions,
using a uniform dielectric constant  (discussed in the following paragraph),
 \begin{equation}
 \begin{aligned}
 V_{\text{coul}} = \frac{e^2}{4\pi\epsilon_0\epsilon_r}\Bigg (&
 \frac{1}{|\mathbf{r}_{e_1}-\mathbf{r}_{e_2}|}
 +\frac{1}{|\mathbf{r}_{h_1}-\mathbf{r}_{h_2}|}\\
 -&\frac{1}{|\mathbf{r}_{e_1}-\mathbf{r}_{h_1}|}
 -\frac{1}{|\mathbf{r}_{e_2}-\mathbf{r}_{h_2}|}\\
 -&\frac{1}{|\mathbf{r}_{e_2}-\mathbf{r}_{h_1}|}
 -\frac{1}{|\mathbf{r}_{e_1}-\mathbf{r}_{h_2}|}
\Bigg),
\end{aligned}
\end{equation}
and there are separate electron and hole confining potentials, $V_e$ and $V_h$, arising from the band edges in the core shell nanocrystal.
\begin{equation}
V_{\text{dot}}=
 V_e(\textbf{r}_{e_1}) 
+ V_e(\textbf{r}_{e_2}) 
+ V_h(\textbf{r}_{h_1}) 
+ V_h(\textbf{r}_{h_2}).
\end{equation}
The exciton Hamiltonian is a simple reduction from this form.
This Hamiltonian treats carrier propagation in the nanocrystal systems within 
the single-band effective mass approximation, and we model the heterointerfaces as 
step-like potentials in the conduction and valence bands. This simplification of the 
semiconductor band structure is the main limitation of the present model, and 
one might expect more accurate predictions to be achieved using a multi-band description. 
The latter would be expected to increase somewhat the degree of correlation due to 
Coulomb interactions, as it is known to reduce the energy spacing between quantum 
confined valence band states as a result of mixing effects \cite{Tyrrell11}.

Our model uses a finite potential barrier for the surrounding matrix, and we assume a 
uniform dielectric constant throughout. We do not consider dielectric polarization effects 
in this work\cite{Bolcatto01}. To check that this approach does not introduce substantial errors, we 
performed a series of calculations that included the dielectric self-energy of the carriers, 
by adding the self-energy potential to the confinement potential before running the 
PI-QMC algorithm. The self-energy potential as a function of radial coordinate was 
calculated according to the work in Refs.~\onlinecite{Tyrrell11}
and \onlinecite{Bolcatto01}. Our results confirmed 
that although inclusion of this potential affected the single particle energy levels 
significantly, it left the exciton binding and exciton-exciton interaction energies 
unchanged, in agreement with previous work \cite{Piryatinski07}.
  
We consider several different types of nanocrystals, initially Type-II core/shell structures for both electron/hole and hole/electron confinement. Later we address some suggestions about the possibility of large X-X interactions in inverted Type-I structures. Finally we present some results for core/rod nanocrystal structures, in which there is currently significant interest due to their excellent as-grown uniformity and very high quantum yields \cite{Talapin07}.

\section{Methods}
The PI-QMC method allows essentially exact treatment of correlation energy, with no basis-set or variational bias.
It is based around a stochastic sampling of the many 
body thermal density matrix,
\begin{eqnarray}
 &&\rho(\textbf{R},\textbf{R}';\beta) = \frac{1}{Z} \langle \textbf{R} \mid e^{-\beta H}\mid \textbf{R}' \rangle,
\end{eqnarray}
where $\beta=1/k_BT$ is the inverse temperature, 
$\textbf{R} = (\textbf{r}_{e_1},\dots,\textbf{r}_{h_2})$ are the particle coordinates, and
the partition function,  $Z = \operatorname{tr}( e^{-\beta H})$, normalizes the density matrix.

Averages of any physical observables $\mathcal{O}$ can then be calculated by,
\begin{equation}
\langle \hat{\mathcal{O}} \rangle = \int d\textbf{R} ~d\textbf{R}' \,
\rho(\textbf{R},\textbf{R}';\beta) \langle \textbf{R}|\hat{\mathcal{O}}| \textbf{R}' \rangle.
\label{operator}
\end{equation}
In the imaginary time path integral method,
the thermal density matrix is expanded, using the primitive approximation \cite{Ceperley95}, into $N$ slices,
\begin{equation}
\begin{aligned}
 &\rho(\textbf{R}_0,\textbf{R}_N;\beta) =
 \frac{1}{Z}
 \left(\frac{2\pi \hbar \Delta\tau}{m}\right)^{-3N/2}
 \int d\textbf{R}_1 d\textbf{R}_2 \dots d\textbf{R}_{N-1}\\
& \times \exp \Bigg[-\sum^{N}_{n=1}\Bigg(\frac{(\textbf{R}_{n-1} - \textbf{R}_n)^2}
 {2\hbar\,\Delta\tau/m}
 +\frac{\Delta\tau\;V_{\text{coul}}(\textbf{R}_n)}{\hbar} \\
 & +\frac{\Delta\tau\;V_{\text{dot}}(\textbf{R}_n)}{\hbar} 
 \Bigg)\Bigg],
\end{aligned}
\end{equation}
with $m$ the mass of the path's particle, and with a time step $\Delta \tau = \beta\hbar/N$.
The $1/r$ singularity of the Coulomb interaction requires extra care, so we replace
the $\Delta\tau\;V_{\text{coul}}$ action term with the pair approximation to the Coulomb 
action~\cite{Ceperley95}.
We then use the well known Metropolis Monte Carlo algorithm for the stochastic 
sampling of the discretized thermal density matrix, in which a closed quantum 
path (such that $\textbf{R}_0=\textbf{R}_N$) representative of a particle is 
randomly walked through real space.

The algorithm is run until the required accuracy is reached, which must be high, as the binding energy is the difference between large total energies. Thus these total energies must have a small absolute errors (typically $\pm$0.5\,meV, unless otherwise indicated with error bars) so as not to result in binding energies with large relative errors. The PI-QMC method has several advantages: the use of the thermal density matrix naturally provides finite temperature simulations, where the temperature is a controllable parameter; it requires no basis set information or trial data; and it can treat spatially complicated potentials.\cite{McDonald10} Finally, and most importantly for the work presented here, it treats the many body Coulomb correlations exactly.

\section{Type II C\lowercase{d}T\lowercase{e}/C\lowercase{d}S\lowercase{e}} We first study Type-II CdTe/CdSe quantum dots as described by Oron \textit{et al}.\ \cite{Oron07} who have presented experimental results for a  such core/shell structures,  in which a transition from the binding to anti-binding regime in the X-X interaction energy is clearly visible as the shell thickness is increased. In our model of this CdTe/CdSe (core/shell) quantum dot, the potential minimum for holes lies in the core and for electrons lies in the shell. 
The band gaps for CdTe and CdSe are taken as 1.475\,eV and 1.75\,eV with the CdTe/CdSe valance band offset taken as -0.57\,eV \cite{Singh,Wei98}. The external potential band gap is assumed to be 4.832\,eV with a valence band offset of 1.325\,eV. We take the electron and hole effective masses isotropically as 0.13\,$m_e$ and 0.35\,$m_e$ respectively.\cite{Spring} All our simulations for this dot were carried out at 300\,K. A dielectric constant of 6.65 is used throughout the simulation unit cell.

\begin{figure}[tbp]
	\centering	
\includegraphics{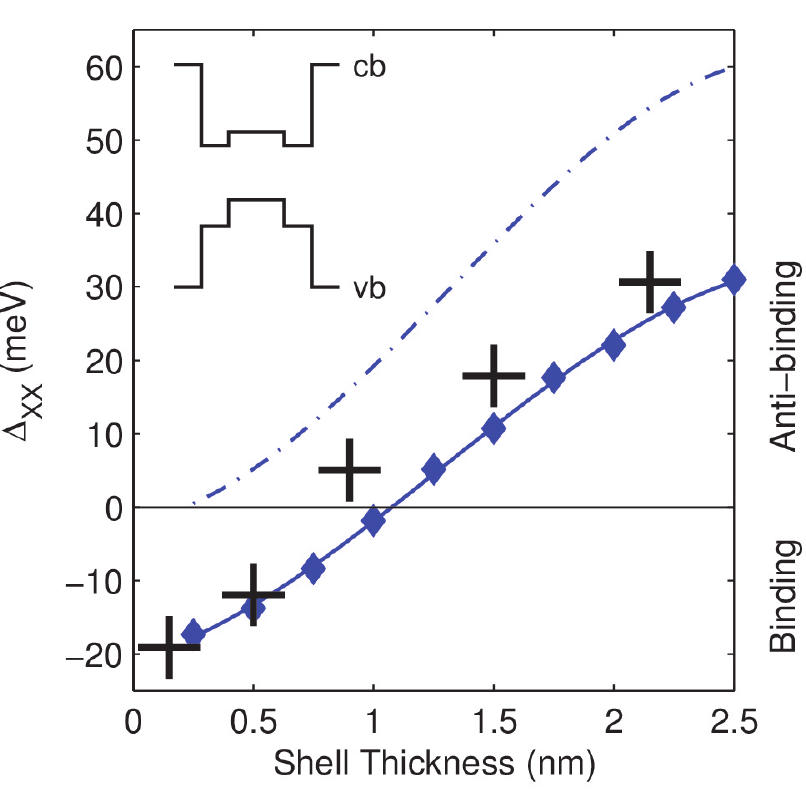}
	\caption{Exciton-exciton interaction energy, $\Delta_{xx}$, versus CdSe shell thickness for a CdTe core diameter of 3.9\,nm.  Experimental data from Oron \textit{et al}.\ \cite{Oron07} are black crosses, with experimental uncertainty approximately given by the symbol size. Blue diamonds show the PI-QMC results, and the line is a guide to the eye. The dashed blue line shows perturbation theory results. Inset shows the radial form of the confinement potential.}
	\label{fig:cdtecdse_result_pert}
\end{figure}
Fig.~\ref{fig:cdtecdse_result_pert} shows the exciton-exciton interaction energy, $\Delta_{xx}$, plotted against shell thickness. Experimental data points are taken from Oron \textit{et al}.\ \cite{Oron07} for a 3.9\,nm diameter core. We see excellent quantitative agreement with these experimental results, demonstrating the importance of a  correct treatment of correlation in calculating the exciton-exciton interaction energies, even in this rather strongly confined nanostructure. 
At the smallest shell thicknesses in Fig.~\ref{fig:cdtecdse_result_pert} the modeled shell thickness is smaller than the lattice spacing and, as in the experiments, reflects an ensemble average over many nanocrystals.  

\begin{figure}[tbp]
	\centering	
\includegraphics{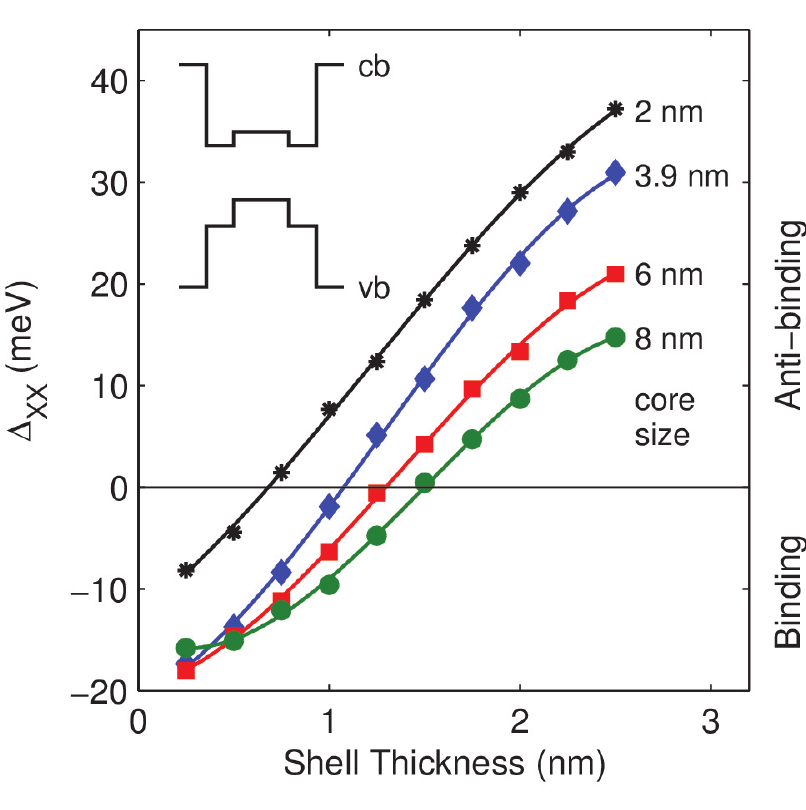}
	\caption{Exciton-exciton interaction energy ($\Delta_{xx}$) versus CdSe shell thickness for various different CdTe core diameters, as indicated. Lines are a guide to the eye. Inset shows the radial form of the confinement potential. }
	\label{fig:cdtecdse_result_XX}
\end{figure}

In Fig.~\ref{fig:cdtecdse_result_XX} we plot $\Delta_{xx}$  against shell thickness for a variety of core diameters. The same trend is evident for all four data sets; for small shell thicknesses the biexciton is strongly bound, with large binding energies. Small shell thicknesses result in a quasi Type-I structure, where the shell is not thick enough to induce localization of the electron, thus the electron is spread across the core, close to the hole, leading to binding of the biexciton. For thicker shells, localization of the electron in the shell begins, and the transition to positive $\Delta_{xx}$  is seen, due to the increased electron-hole seperation and hence a reduced attractive Coulombic interaction energy, whilst the large repulsive hole-hole interaction from the core confined holes remains relatively unchanged.

Core/shell nanocrystals with smaller cores have a smaller binding energy and a larger anti-binding energy, due to the stronger confinement of the holes in the core, giving a larger repulsive element to the binding energies. The smaller the core size, the larger the final anti-binding, as the separation between the holes within the core remains small despite their mutual repulsion.

\begin{figure}[tbp]
	\centering	
\includegraphics{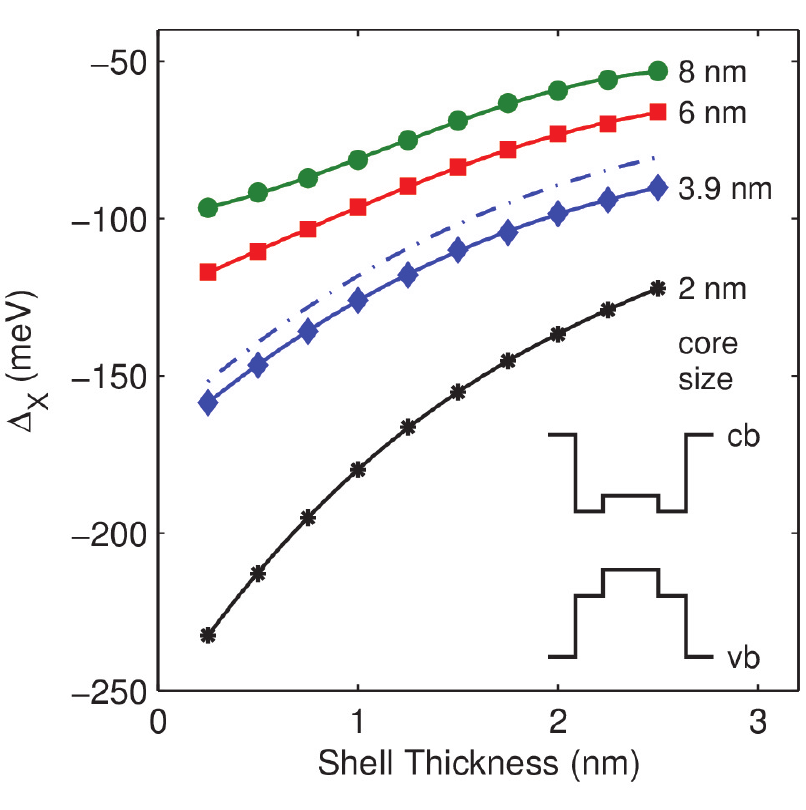}
	\caption{Exciton interaction energies, $\Delta_{x}$, as a function of CdSe shell thickness for various CdTe core diameters, as indicated. Perturbation theory result for 3.9\,nm core are shown as a dashed blue line. Lines are a guide to the eye. Inset shows the radial form of the confinement potential.}
	\label{fig:cdtecdse_result_X}
\end{figure}

We now compare our PI-QMC results to that of first order perturbation theory to assess the role of correlation. Quite good agreement is observed between the exciton interaction energies, $\Delta_{x}$, calculated by the two methods as shown in Fig.~\ref{fig:cdtecdse_result_X}. $\Delta_{x}$ is the energy change attributable to the Coulomb interaction, \textit{i.e.}, the negative of the binding energy. The Coulomb correlation between the electron and hole wavefunctions that are included in the PI-QMC results increase the binding energy by 7-10\,meV compared with the perturbation theory predictions. 
For the biexciton (Fig.~\ref{fig:cdtecdse_result_pert}) the difference in $\Delta_{xx}$ between the two approaches is more marked.
The perturbative approach fails to predict any transition between negative and positive values of $\Delta_{xx}$, and further, over estimates the anti-binding significantly. The correlations included in the PI-QMC approach lead to a reduction in $\Delta_{xx}$ of about 17\,meV in the core-only Type-I structure, and which gradually increases to about 29\,meV for the fully Type-II structure with a 2.5\,nm shell. The increasing importance of the correlations with increasing shell thickness can be attributed to the departure from the strong confinement limit as the electron becomes less localized. 
Calculations performed for a 3.9\,nm core with thicker shells (not shown) reveal that $\Delta_{xx}$ saturates to about 70\,meV at a shell thickness of around 7.5\,nm.

\begin{figure*}[tbp]
	\centering	
\includegraphics{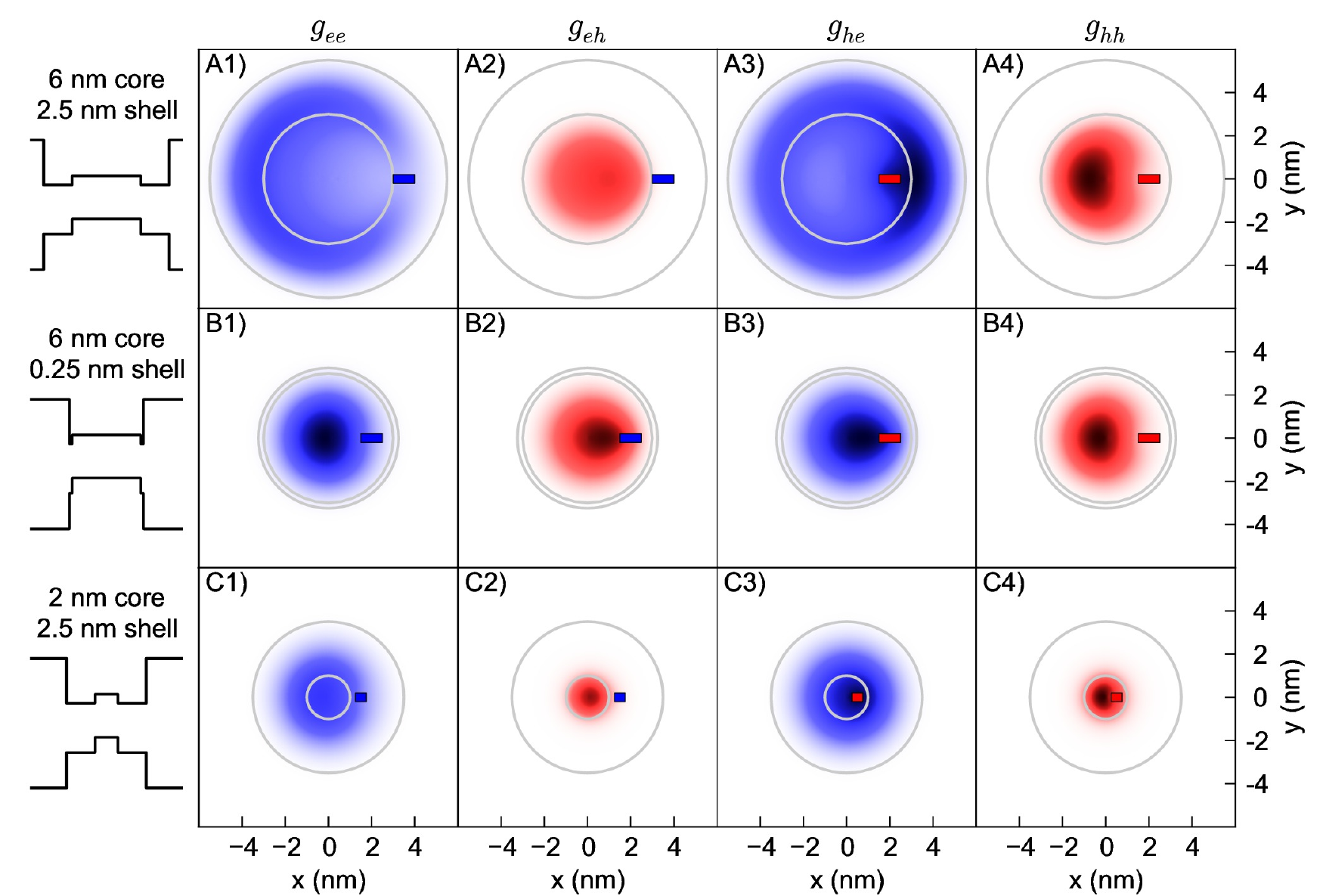}
\caption{Conditional probability densities are shown for a  6\,nm CdTe core diameter and 2.5\,nm CdSe shell thickness in row A, 
for a  6\,nm CdTe core diameter and 0.25\,nm CdSe shell thickness in row B and for a  2\,nm CdTe core diameter and 2.5\,nm CdSe shell thickness in row C.
The radial form of the confinement potential for each is illustrated.
Shown in column 1) is $g_{ee}$, a conditional electron (falling within the blue rectangle) and the resulting electron distribution, 
column 2) shows $g_{eh}$, a conditional electron and resulting hole distribution, 
column 3) shows $g_{he}$, a conditional hole (falling within the red rectangle) and resulting electron distribution and 
column 4) shows $g_{hh}$, a conditional hole and resulting hole distribution. }
	\label{fig:cdtecdse_cond}
\end{figure*}

The role of correlations in the biexcitons can be examined directly through the conditional probability density function for the electrons and holes as shown in Fig.~\ref{fig:cdtecdse_cond}.
 Each sub-panel A1-A4 in Fig.~\ref{fig:cdtecdse_cond} shows a slice of the conditional probability density through the x-y plane of a 6\,nm CdTe core diameter colloidal dot with a 2.5\,nm CdSe thick shell. This geometry places it strongly in the anti-binding regime, as seen in Fig.~\ref{fig:cdtecdse_result_XX}. 
The small rectangle in each panel indicates the location of a small volume, $\cal R$, which subtends a solid angle  of 0.5 degrees between the two radii at either end of the rectangle. (The rectangles are drawn wider than 0.5 degrees for illustrative purposes.) 

The conditional probability density in each panel along the row (1)-4)) is defined by 
\begin{equation}
g_{ij}(\textbf{r}) = \int_{\cal R} \langle n_{i}(\textbf{r}')~ n_{j}(\textbf{r}) \rangle~ d^{3}\textbf{r}',
\end{equation}
where $i$ denotes the particle that must fall into the region $\cal R $ in order for the location of particle $j$ to be sampled.
Hence plotted in sub-panel A1 is the probability of observing an electron at a given position if a first electron is found in the region $\cal R $.
Similarly Fig.~\ref{fig:cdtecdse_cond} A2 shows the pair correlation density for holes when an electron is present inside the rectangle. Fig.~\ref{fig:cdtecdse_cond} A3 and A4 are analogous plots for when a hole is located in the region $\cal R$. 
In Fig.~\ref{fig:cdtecdse_cond} A1 it can be seen that the electrons repel one another and sit on opposite sides of the dot; the same behavior is seen between the two holes in A4 with the holes sitting on opposite sides of the core. In A2 and A3 opposite charges can be seen correlated and being attracted towards the opposite charge located in the region $\cal R$. It is clear that in addition to the radial correlations, angular correlations also play a role in reducing the electron-electron and hole-hole interactions to further reduce the total Coulomb energy in the system.

We contrast this with the case of the bound biexciton, for a 6\,nm CdTe core diameter and 0.25\,nm CdSe shell thickness shown in the second row of Fig.~\ref{fig:cdtecdse_cond}, again similar  electron-electron and hole-hole repulsion is seen in B1 and B4. However, the strong quantum confinement of the electrons and holes keeps them well confined to the core, leading to the strong electron-hole interactions seen in B2 and B3. This  results in a much increased overlap between the electron and holes charge densities and this increase in the attractive electron-hole interaction gives rise to a bound biexciton. The correlations indicate a state in which the electron and hole motions are strongly overlapping. 

In a dot with a small 2\,nm diameter CdTe core and 2.5\,nm CdSe shell thickness, we see a significantly increased anti-binding. In the final row of Fig.~\ref{fig:cdtecdse_cond} C2 and C4 shows the hole being  strongly confined in the small core, having little room to avoid the other hole, resulting in large repulsive terms contributing to the anti-binding. The electrons, shown in Fig.~\ref{fig:cdtecdse_cond} C1 and C4 are more spread  out relative to the hole when compared to the binding example of Fig.~\ref{fig:cdtecdse_cond}, albeit with a smaller overall volume.

\section{Type II C\lowercase{d}S/Z\lowercase{n}S\lowercase{e}} There have been several publications focused on a similar Type-II colloidal core/shell nanostructure \cite{Klimov07, Nanda2007}, namely CdS/ZnSe. The band gaps and offsets for this particular combination result in an opposite potential profile compared to a CdTe/CdSe nano-crystal, with the hole potential minimum in the shell, and the electron minimum in the core. 
A similar transition from binding to anti-binding of the biexciton is expected in this Type-II structure.

\begin{figure}[tbp]
	\centering	
\includegraphics{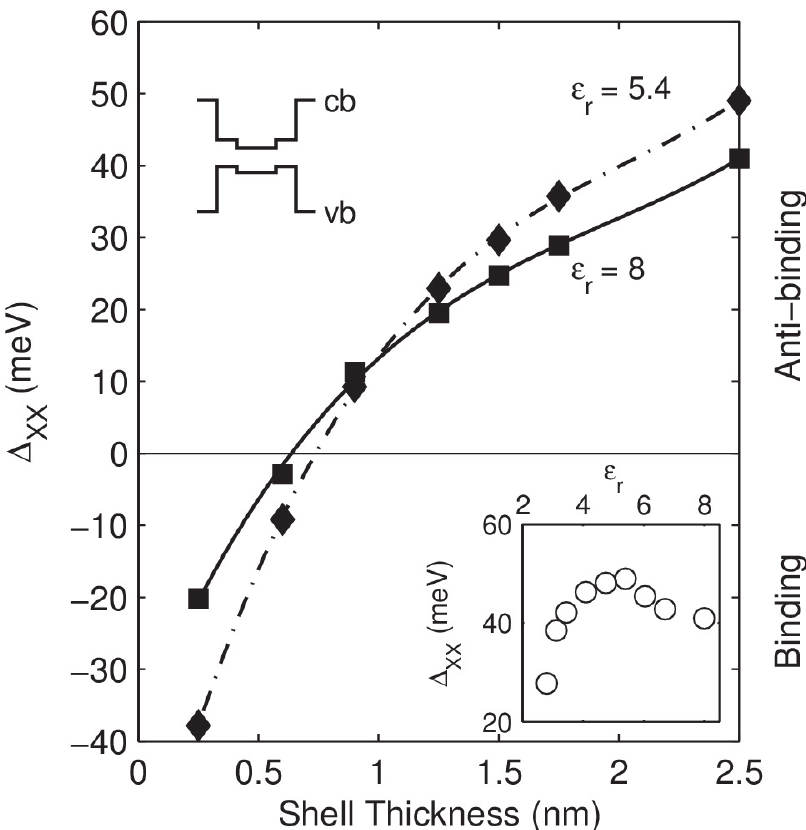}
	\caption{Calculated exciton-exciton interaction energy, $\Delta_{xx}$, against shell thickness for CdS/ZnSe Type-II colloidal nanocrystal, with the radial form of the confinement potential shown in upper inset. Core diameter of 3.1\,nm, with dielectric constant of $\epsilon_r$=8 (squares with solid line) and for a  weaker dielectric constant of $\epsilon_r$=5.4 (diamonds and dashed line). Lower inset shows $\Delta_{xx}$ against
$\epsilon_r$, for a fixed shell thickness of 2.5\,nm. }
	\label{fig:cdsznse_result}
\end{figure}

Particularly noteworthy is the extremely strong measured biexciton anti-binding of $\sim$100\,meV which led to a demonstration of single exciton optical gain\cite{Klimov07}. Perturbation theory results provide some support for these values when low values of the dielectric constants are employed\cite{Piryatinski07}. 
Again in this case perturbative calculations overestimate the anti-binding properties and miss the transition from binding to anti-binding.
For our model system we use a CdS core with a bulk band gap of 2.485\,eV and a ZnSe shell with a bulk band gap of 2.720\,eV, and a conduction band offset of 0.795\,eV as shown in Fig.~\ref{fig:cdsznse_result} (inset) \cite{Klimov07}. Electron and hole masses of 0.2\,$m_e$ and 0.6\,$m_e$ respectively are used.  We perform all simulations at 300\,K.

In Fig.~\ref{fig:cdsznse_result} we see a similar transition as in the CdSe/CdTe Type II structures, with a transition from binding to anti-binding at a shell thickness of around 0.7\,nm, only slightly more than one monolayer shell coverage.
The anti-binding is significantly less than the $\sim$100\,meV measured in Ref. ~\onlinecite{Klimov07}, the uncertainty in dielectric constant, masses and core/shell sizes and shapes may account for some of this difference, but is unlikely to increase the maximum to this level within the shell thickness range described here.
Fig.~\ref{fig:cdsznse_result} also shows the effect of a decreased dielectric constant from $\epsilon_r=8$  to $\epsilon_r=5.4$, and this is found to give both larger binding and anti-binding values. However, a maxima is found at around $\epsilon_r=5.4$, with lower dielectric constants than this leading to weaker anti-binding or even binding behavior, as shown in Fig.~\ref{fig:cdsznse_result} (lower inset). Lower dielectric constants lead to an increase in the coulomb interaction strength, and in the limit of small dielectric constants the four attractive terms can dominate, causing binding. In the limit of large dielectric constants, the coulomb interaction strength tends to zero leading to a $\Delta_{XX}$ of zero. The maxima lies somewhere in between these two limits, here at approximately $\epsilon_r=5.4$.

\begin{figure}[tbp]
	\centering	
\includegraphics{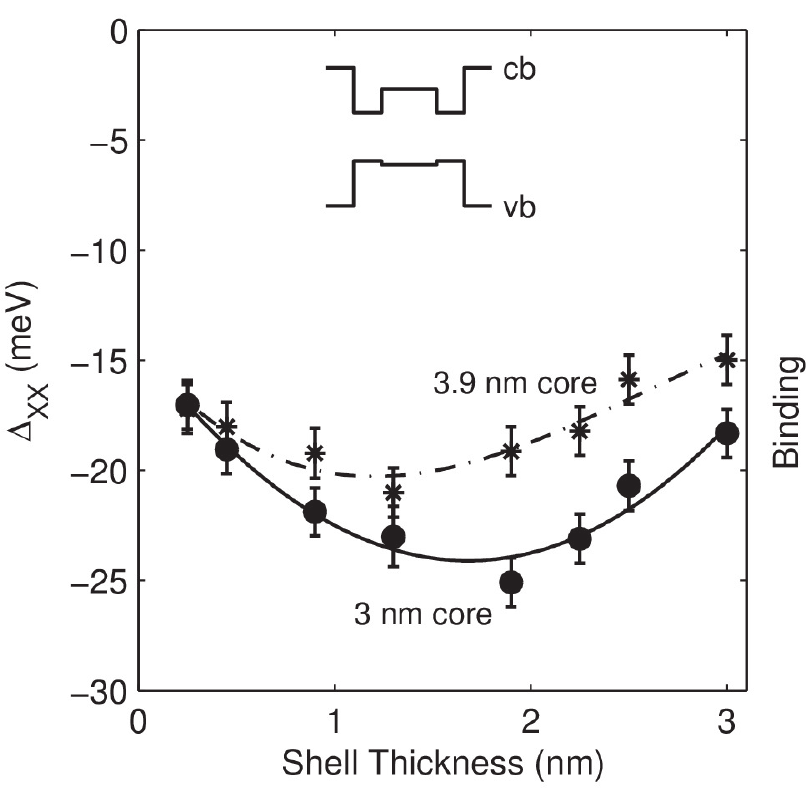}
	\caption{Exciton-exciton interaction energy, $\Delta_{xx}$,  plotted against increasing shell thickness for ZnSe/CdSe inverted Type-I colloidal nanocystal with 3.0\,nm (circles with solid line) and 3.9\,nm (stars with dashed line) core diameters. The inset shows the radial form of the confinement potential and  lines are a guide for the eye. Simulations are performed at 300\,K.}
	\label{fig:inverted_result}
\end{figure}

\begin{figure*}[tbp]
	\centering	
\includegraphics{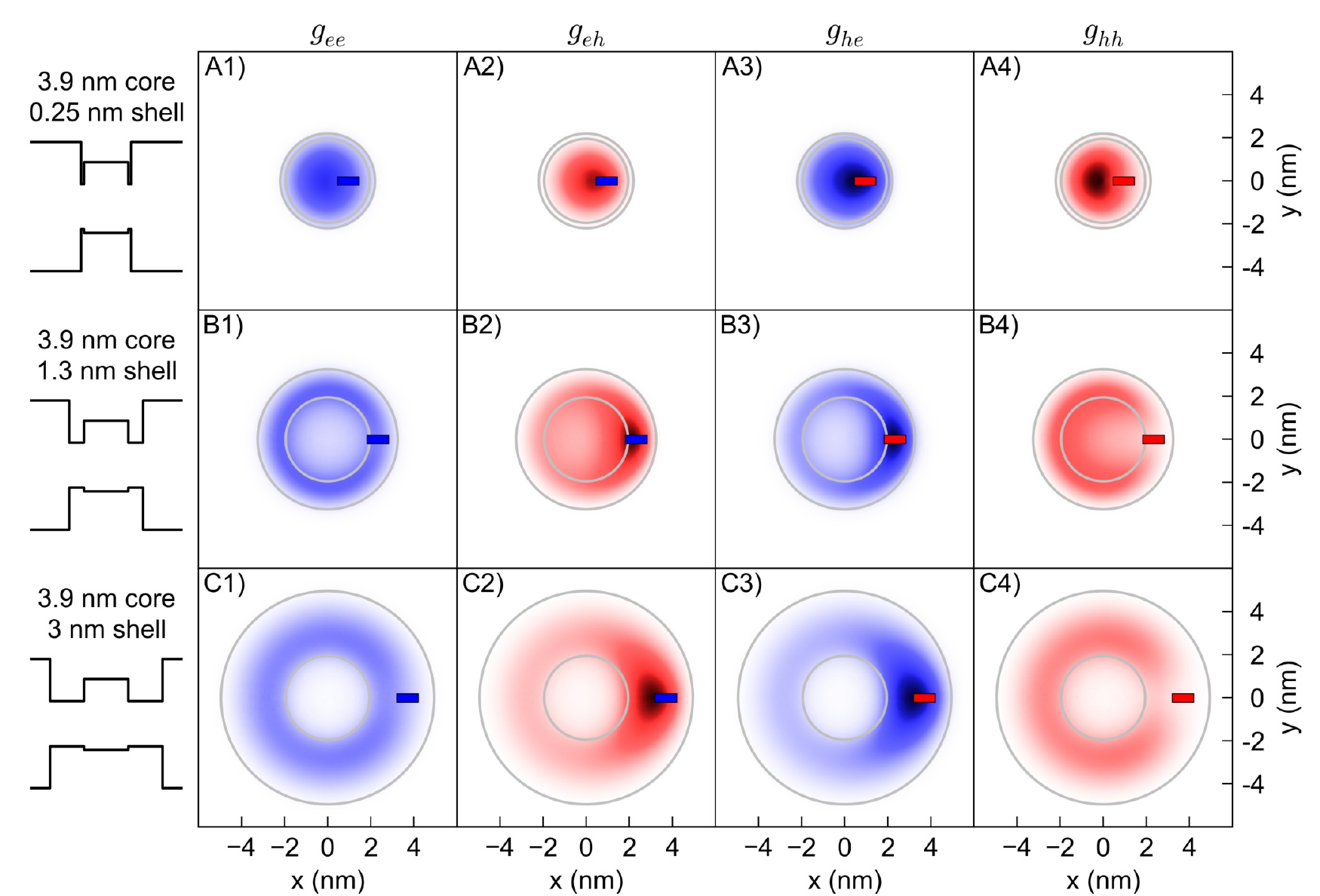}
\caption{Conditional probability densities are shown for a  3.9\,nm ZnSe core diameter and 0.25\,nm CdSe shell thickness in row A.
For a 3.9\,nm CdTe core diameter and 1.3\,nm CdSe shell thickness in row B, and for a 3.9\,nm ZnSe core diameter and 3\,nm CdSe shell thickness in row C.
The radial form of the confinement potential for each is illustrated.
Shown in column 1) is $g_{ee}$, a conditional electron (falling within the blue rectangle) and the resulting electron distribution, 
column 2) shows $g_{eh}$, a conditional electron and resulting hole distribution, 
column 3) shows $g_{he}$, a conditional hole (falling within the red rectangle) and resulting electron distribution and 
column 4) shows $g_{hh}$, a conditional hole and resulting hole distribution. }
	\label{fig:inver_cond}
\end{figure*}

\section{Inverted Type-I Z\lowercase{n}S\lowercase{e}/C\lowercase{d}S\lowercase{e}} It has further been suggested \cite{Ivanov04, Nanda2006} that a similar transition in the biexciton binding may be seen in an inverted Type-I structure, such as in the ZnSe/CdSe core/shell structure, where the electron and hole both have potential minima inside the shell. 
The band offset for the valence band is small (0.14\,eV), and for small shell thickness the hole may delocalize across the entire structure. By comparison the conduction band offset is much larger (0.86\,eV)\cite{Ivanov04}. Therefore a scenario where the hole is delocalized across the structure, and the electron confined to the shell, may lead to a quasi Type-II structure\cite{Nanda2006}. As the shell thickness is increased we would therefore expect in a simple picture to go from binding where both electron and hole are localized in the core, to anti-binding as described in the quasi Type-II scenario, and then back to binding, where again the electron and hole are localized in the shell.

We use our PI-QMC calculations to go beyond this simple picture and include correlations as shown in Fig.~\ref{fig:inverted_result}. 
Compared to the simple discussion above however, we see in fact  the opposite trend. 
Due to the correlations we find that the biexciton is bound for all shell thicknesses.

As shown in Fig.~\ref{fig:inver_cond} sub-panel A1-A4, for a  3.9\,nm ZnSe core and a thin CdSe shell of 0.25\,nm,  both the electrons and holes are confined to the core of the dot, and a bound biexciton forms.
As the shell thickness is increased to 1.3\,nm (Fig.~\ref{fig:inver_cond} sub-panel B1-B4), the electron and hole (more slowly, due to the lower valence band offset) become more confined to the shell. 
In Fig.~\ref{fig:inver_cond} B1 the electrons are seen to  repel each other, however, not so strongly as to localize on the opposite sides of the dot as in the previously described Type II anti-binding cases. This same feature is evident in the hole-hole repulsion shown in B4. At the same time B2 and B3 show excitons forming. 
The bound exciton seen in B2 can be compared with the conditional density in B4 where  the correlation hole matches very closely with the shape of B2, with the same feature visible for the electrons in B1 and B3. This case  can be understood in that there is strong Coulomb localization of electrons around the heavier, less well confined holes. Thus the presence of a hole in the close vicinity of the electron renders the pair effectively neutral and mitigates the inter-electron repulsion.  The mechanism for the bound biexciton in this system is that of two interacting excitons with a weak mutual attraction.
As the shell thickness increases the increased volume of the dot allows like charges to spatially separate more and results in a slight increase in binding.

The binding decreases again for larger shell thicknesses, as the electron and hole become very well confined to the shell and like particles are forced closer together, increasing the repulsion felt, and reducing the binding energy.

Such complex interplay between particles is handled well in the PI-QMC calculation, and as shown, can lead to results different to those produced by simple models.

\begin{figure}[tbp]
	\centering	
\includegraphics{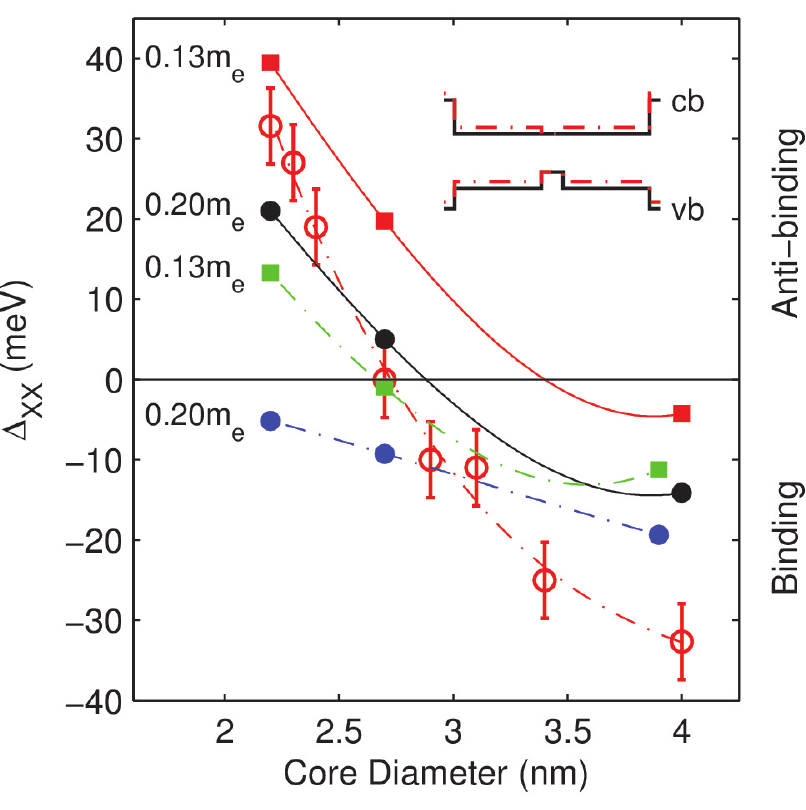}
	\caption{The effect of various parameters on the binding to anti-binding regime transition for CdS/CdSe rod/core nanocrystal. Red circles with error bars are experimental data taken from Sitt \textit{et al}.\ \cite{Sitt09}, along with polynomial fit to experimental data (red dashed line). Electron masses used for each dataset are indicated, solid lines indicate a 0\,eV conduction band offset. Dashed lines indicated a conduction band offset of 0.3\,eV.  All simulations are with $\epsilon_r$=8. Inset shows form of confinement potential, black solid line indicates potential with 0\,eV conduction band offset, red dashed line shows potential with 0.3\,eV conduction band offset. Lines are a guide to the eye.}
	\label{fig:rod_results_various}
\end{figure}

\begin{figure}[tbp]
	\centering	
\includegraphics{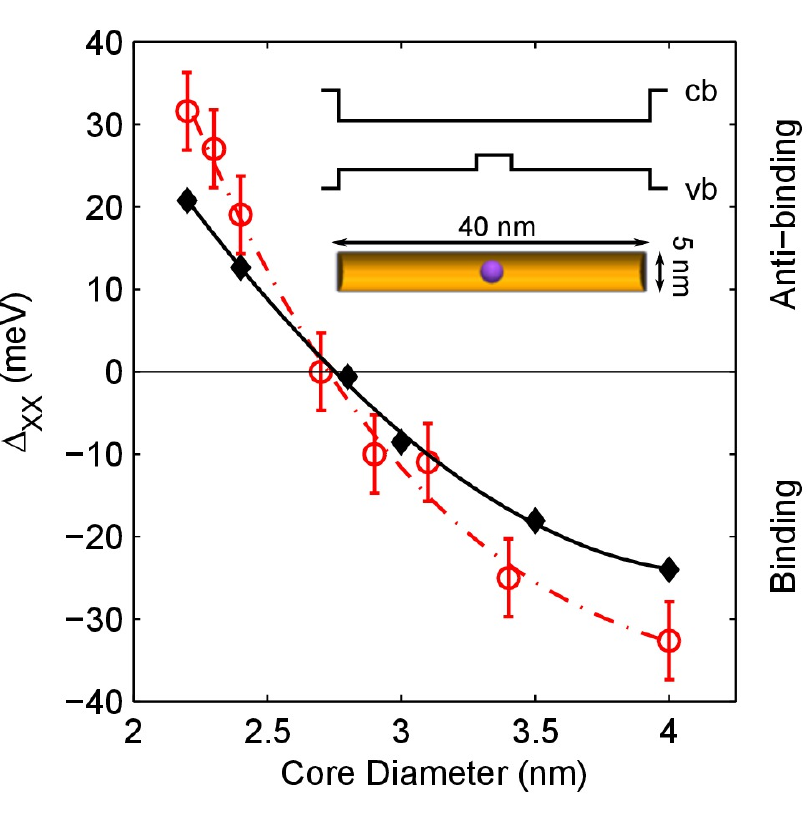}
	\caption{Exciton-exciton interaction energy, $\Delta_{xx}$, against core diameter, for a nanocrystal with a CdS rod and CdSe core. Closest fit of PI-QMC simulation (black diamonds) to experimental data from Sitt \textit{et al}.\  \cite{Sitt09} (red circles, dashed red line is polynomial fit to experimental data points) - is with 0\,eV conduction band offset and 0.2\,$m_e$ electron mass, with $\epsilon_r$=5.785. Inset shows example confinement potential with 0\,eV conduction band offset and position of core in rod.}
	\label{fig:rod_results}
\end{figure}

\section{C\lowercase{d}S\lowercase{e}/C\lowercase{d}S dot/rod} There has also been substantial  discussion about colloidal dot/rod structures, in which a colloidal nanocrystal is embedded inside a nanorod. Sitt \textit{et al}.\ \cite{Sitt09} reported a similar biexciton binding/anti-binding transition as in Type-II nanocrystals. The CdSe/CdS dot/rod structure, has been suggested to have either Type-I, or quasi Type-II confinement. The debate arises from  the current uncertainty regarding the correct conduction band offsets. Experimental results however see a strong transition from biexciton binding to anti-binding, indicating at least some Type-II characteristic behavior. 
We base our model system on that used by Sitt \textit{et al}.\ \cite{Sitt09}. A rod of fixed length 40\,nm is chosen, and the dot size systematically varied. The CdSe core is taken to have a band gap of 1.75\,eV whilst the CdS rod has a band gap of 2.5\,eV, and a fixed dielectric constant of 8 is initially used. We use 0.3\,eV and 0\,eV conduction band offsets to investigate the effect of these on the exciton-exciton interaction energy, $\Delta_{xx}$. We take the hole mass to be that of the CdS core, 0.4\,$m_e$, and the electron mass to be either 0.13\,$m_e$ or 0.2\,$m_e$. We perform all simulations at 300\,K.

We begin with the core in the middle of the rod. In our PI-QMC calculations, Fig.~\ref{fig:rod_results_various}, we find that the large offset of 0.3\,eV with a heavy electron mass, 0.2\,$m_e$, confines the electron strongly to the core, resulting in an always-bound biexciton, resembling Type-I behavior. A lighter mass of electron, 0.13\,$m_e$, allows it to delocalize further outside the dot and along the rod, showing quasi Type-II behavior. Similar behavior is seen for a flat offset of 0\,eV with the heavier electron mass. A flat offset paired with a light electron mass gives a nearly continuously anti-bound biexciton, which puts it strongly in the Type-II regime. It is clear that the biexciton binding transition in the rod is particularly sensitive to the choice of the band offset and effective mass parameters.

In order to match closely the experimental data, we use a dielectric constant of the average of CdS and CdSe, of 5.785, with a 0\,eV conduction band offset and a heavier electron mass of 0.2\,$m_e$. As we see in Fig.~\ref{fig:rod_results}, this gives excellent agreement with experimental results. Interestingly, moving the dot nearer to the end of the rod was found to have very  little impact on the transition.

\begin{figure}[tbp]
	\centering	
\includegraphics{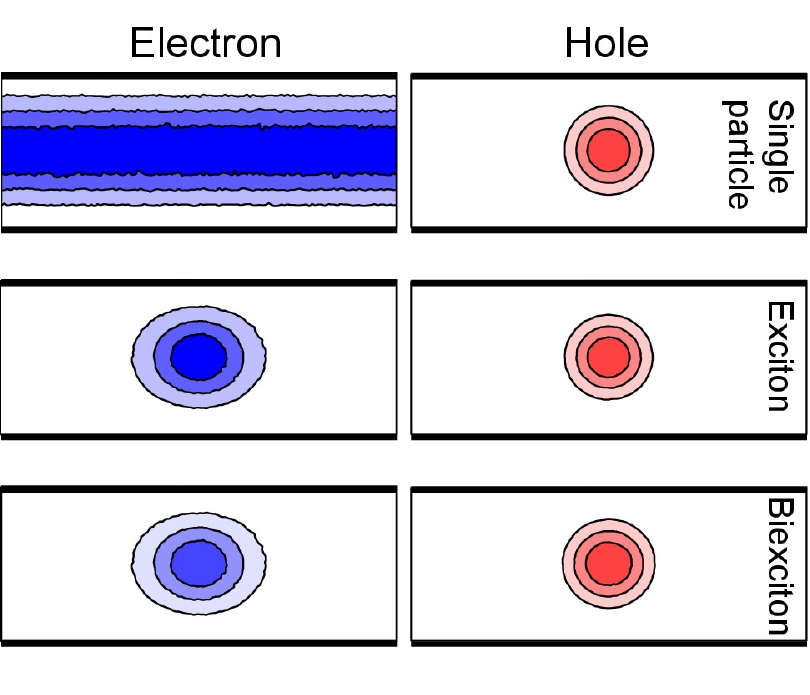}
\caption{Probability densities showing the role of correlation due to the Coulombic interaction  in CdS/CdSe rod/core structure, here with a core diameter of 4\,nm. The single particle electron is highly delocalized along rod, the attractive Coulomb potential then strongly localizes the electron to the hole, as seen in the exciton and biexciton densities. Black lines show the width of the rod (5\,nm), with the central 12.8\,nm of the rod shown out of the total 40\,nm length.}
\label{fig:rod_coloumb}
\end{figure}

We note that the normal perturbative method for calculating $\Delta_{xx}$ is a particularly poor approximation for the dot/rod structures. The single particle electron and hole  densities for the core/rod structures are shown in Fig.~\ref{fig:rod_coloumb}. As expected, we see a strongly delocalized electron and a localized hole. However, correlation effects in these rod structures are striking. Including  the Coulomb interaction in the path integral calculation, we can see the electron density is much more strongly confined towards the core by its attraction to the hole. This has the effect of strongly increasing the binding aspect of the biexciton transition. 
Perturbative methods neglect this correlation-enhanced binding; hence, using only perturbative methods to infer the conduction band offset from experimental data can be misleading.

\section{Conclusions} 
We have presented for the first time theoretical calculations for colloidal quantum dots, which 
systematically show the exciton-exciton interaction energy undergoing a transition from the strongly binding regime to the strongly anti-binding regime.
Our results illustrate the significance of coulomb correlations in determining biexciton binding energies, and show excellent agreement with experimental data for CdTe/CdSe Type-II nanocrystals over a range of quantum dot sizes. In CdS/ZnSe Type-II nanocrystals, we find our results do not agree with the large anti-binding values previously found, with the effects of higher or lower dielectric constants unable to account for the discrepancy --- raising new questions as to the anti-binding mechanism in these dots.
We also demonstrate that for the inverted Type-I nanocrystal system, coulomb correlations can lead to a qualitatively different mechanism of binding and anti-binding when compared to models which exclude these correlations; due to this our results always show bound biexcitons.
Lastly, we provide insight into novel dot/rod structures in which correlations are very strong and perturbative methods are particularly inaccurate. These methods will be of importance in assessing the potential of particular nanocrystal structures as single exciton lasing media. 
\begin{acknowledgements}
P.G.M and E.J.T acknowledge funding from EPSRC. J.M.S acknowledges additional financial support from Hewlett Packard Ltd. Computational resources were provided by Heriot-Watt University and the Arizona State University Center for Advanced Computing (A2C2).
\end{acknowledgements}

\bibliography{references.bib}

\end{document}